\documentclass[apj]{emulateapj}
\usepackage{apjfonts, xspace}
\shorttitle{Synchrotron Constraints on a Hybrid Galactic Wind}
\shortauthors{Everett, Schiller \& Zweibel}

\newcommand{\pOne}{Paper\ I\xspace}
\newcommand{\degrees}{\ensuremath{^\circ}}

\newcommand{\pc}{\ensuremath{\rm\,pc}\xspace}
\newcommand{\kpc}{\ensuremath{\rm\,kpc}\xspace}

\newcommand{\kms}{\ensuremath{\rm\,km~s^{-1}}\xspace}

\newcommand{\ergsPerSec}{\ensuremath{\rm\,erg\,s^{-1}}\xspace}
\newcommand{\mhz}{\ensuremath{\rm\,MHz}\xspace}

\newcommand{\kev}{\ensuremath{\rm\,ke\hspace{-0.12ex}V}\xspace}
\newcommand{\gev}{\ensuremath{\rm\,Ge\hspace{-0.12ex}V}\xspace}
\newcommand{\muG}{\ensuremath{\mu\rm G}\xspace}

\newcommand{\K}{\ensuremath{\rm\,K}\xspace}

\newcommand{\alfven}{{Alfv\'en}\xspace}

\begin{document}

\bibliographystyle{apj}

\title{Synchrotron Constraints on a Hybrid Cosmic-Ray and
  Thermally-Driven Galactic Wind}

\author{by John E. Everett\altaffilmark{1,2,3}, 
  Quintin G. Schiller\altaffilmark{2},
  Ellen G. Zweibel\altaffilmark{1,2,3}}
\altaffiltext{1}{University of Wisconsin--Madison, Department of Astronomy}
\altaffiltext{2}{University of Wisconsin--Madison, Department of Physics}
\altaffiltext{3}{Center for Magnetic Self-Organization in Laboratory
  and Astrophysical Plasmas}
\email{everett@physics.wisc.edu}

\begin{abstract}
Cosmic rays and magnetic fields can substantially impact the launching
of large-scale galactic winds.  Many researchers have investigated the
role of cosmic rays; our group previously showed that a cosmic-ray and
thermally-driven wind could explain soft X-ray emission towards the
center of the Galaxy.  In this paper, we calculate the synchrotron
emission from our original wind model and compare it to observations;
the synchrotron data shows that earlier assumptions about the
launching conditions of the wind must be changed: we are required to
improve that earlier model by restricting the launching region to the
domain of the inner ``Molecular Ring'', and by decreasing the magnetic
field strength from the previously assumed maximum strength.  With
these physically-motived modifications, we find that a wind model can
fit both the radio synchrotron and the X-ray emission, although that
model is required to have a higher gas pressure and density than the
previous model in order to reproduce the observed X-ray emission
measure within the smaller `footprint'.  The drop in magnetic field
also decreases the effect of cosmic-ray heating, requiring a higher
temperature at the base of the wind than the previous model.
\end{abstract}


\section{Introduction}\label{Intro}

The possibility of thermal-pressure driven outflows on galactic scales
has been considered for some time \citep[see, e.g.,][and references
therein]{VeCeBH2005}.  Somewhat less well-known, however, is the
possibility that cosmic rays and magnetic fields can also help drive
outflows: if, as commonly assumed and even sometimes observed,
magnetic-field pressure and cosmic-ray pressure are in approximate
equipartition with the hot-gas thermal pressure within galactic disks
\citep[e.g.,][]{Duric1990, Pohl1993, Heiles1996, ZweibelHeiles1997,
Webber1998, Beck2001, Cox2005}, these sources of energy will be
important in large-scale outflows in more quiescent galaxies, such as
perhaps our Milky Way.  Understanding and constraining the dynamics
and observational constraints on such kpc-scale galactic outflows is
the main aim of the present paper.

The possibility of a kpc-scale outflow from the Milky Way has been
studied for over three decades \citep[e.g.,][]{Ipavich1975, BVM1987,
BrMcVo1991, BrMcVo1993, BlHaCo03, Breitschwerdt2003, VeCeBH2005},
including work on the observational effects of such an outflow
\citep[e.g.,][]{LercheSchlickeiser1982, JokipiiMorfill1987,
ReichReich1988a, Pohl1990, Bloemen1991, BloemenEtAl1993,
BreitschwerdtSchmutzler1994, Zirakashvili1996, PtuskinEtAl1997,
BrDoVo2002}.  \citet{BlHaCo03} showed possible evidence for such an
outflow in X-ray, infrared, and radio observations, along with models
of a bipolar structure.  In \citet[][hereafter
\pOne]{EverettEtAl2008}, we continued this work by building a wind
model powered by both cosmic-ray pressure and thermal-gas pressure.
In such winds, cosmic rays yield some of their energy to \alfven waves
on the magnetic field via the ``Streaming Instability''
\citep[][]{KulsrudPearce1969} which are then damped to deliver energy
and momentum to the thermal gas.  We applied that model to
observations of excess diffuse X-ray emission towards the inner Galaxy
\citep{SnowdenEtAl1995, SnowdenEtAl1997, AlmyEtAl2000}.  \pOne built
especially on \citet{BrMcVo1991}, but with the following
modifications: the wind is launched from a restricted range of
Galactocentric radii (an annulus in the inner Galaxy), originates at
the Galactic midplane with a vertical magnetic field, and (as
indicated by our analytical work) the cosmic-ray generated \alfven
waves are completely damped, subsequently heating the gas.  This model
fit the longitude-averaged X-ray emission observed with \textit{ROSAT}
near 0.65\kev and 0.85\kev with $\chi^2$ values a factor of two
smaller than the previous static model \citep{AlmyEtAl2000}, while
simultaneously yielding pressure and density parameters approximately
equal to those already inferred for hot gas near this region in the
inner Milky Way \citep[c.f.,][hereafter, F01]{Ferriere2001}.

The above initial model (reviewed in \S\ref{windReview}) was derived
solely from a fit to the soft X-ray emission and made several
assumptions that bear continued examination.  In this paper, we test
the wind model further by comparing calculations of the synchrotron
emission from the wind to 408-MHz all-sky surveys towards the center
of the Galaxy: radio-synchrotron observations are an important
constraint on a cosmic-ray and thermal-gas pressure-driven outflow,
given the dependence on both the cosmic-ray pressure (albeit through
the uncertain and energy-dependent ratio of cosmic-ray electrons to
protons) and the magnetic field strength, both of which are crucial in
our model.  Many groups have worked to understand the distribution of
synchrotron emission in the halo \citep[F01;][]{BeuermannEtAl1985,
ReichReich1988a, ReichReich1988b, BroadbentEtAl1990, StrongEtAl2000,
ReichEtAl2004, SunEtAl2008, OliveriaCostaEtAl2008, ReichReich2008,
WaelkensEtAl2009}, but more work is needed to understand the role of
cosmic-ray advection into the halo via a wind and the subsequent
cooling of cosmic-ray electrons via inverse-Compton and synchrotron
cooling \citep[although we note that the cosmic-ray propagation code
GALPROP has an approximate Galaxy-wide wind model with a constant
velocity gradient; see, e.g.,][and references
therein]{PorterEtAl2008}.

In this work, we build a model of the synchrotron emission of a
large-scale wind launched from an annulus in the inner Galaxy, taking
into account synchrotron and inverse-Compton cooling and the full
velocity gradient in the wind; this work is detailed in
\S\ref{synchEmission}.  We also present predictions for the
synchrotron spectral index, but note the possibly strong importance of
foreground synchrotron emission.  By comparing the observed
synchrotron emission to that calculated for the model of \pOne, the
observations can then be used to examine two assumptions made in
\pOne: the radial range on the Galactic disk where the wind is
launched and the magnetic field strength.

The calculated synchrotron emission will show that the wind model of
\pOne overpredicts the observed radio emission; the observed
synchrotron emission is therefore a strong constraint on the wind
model.  As outlined above, we then consider if it is possible for a
modified wind model to fit the observations; these possibilities are
explained in \S\ref{newModelSetup}. First, the wind may have a thinner
footprint than the $\Delta R = 3\kpc$ width used in \pOne
(\S\ref{thinnerWind}); this is suggested by observations of the
``5\kpc Molecular Ring'' \citep[][B. Benjamin, private
communication]{JacksonEtAl2006}.  Second, the magnetic field strength
within the wind might be lower than the value of $7.8\muG$ used in
\pOne (\S\ref{lowerBField}).  Third, the cosmic-ray electron-to-proton
ratio may be different towards the center of the Galaxy, although we
find it unnecessary to invoke this, given other, more physically
motivated and astrophysically constrained modifications to the model
(\S\ref{initCReDistribution}).  (We also modify the $\chi^2$ fit
parameter to avoid biases due to assumptions about X-ray absorption in
\S\ref{truncChi2}.)  Then, in \S\ref{improvedWindModel}, we present
the best-fit wind model that addresses both the soft X-ray and
synchrotron observations, and compare this model to that of \pOne.
Finally, we discuss the results in \S\ref{results}.

\section{The Hybrid Cosmic-Ray and Thermally-Driven Wind Model}\label{windReview}

Before we outline our model of synchrotron emission from the wind, we
first review the wind model from \pOne.  This model was motivated by
observations of the inner Milky Way, where an excess of $T \sim 3
\times 10^6$\K, X-ray emitting gas has been observed with a
scaleheight of $\sim$2\,kpc \citep{SnowdenEtAl1995, SnowdenEtAl1997,
ParkEtAl1997, ParkEtAl1998, AlmyEtAl2000}.  Some attempts had been
made \citep{SnowdenEtAl1997, AlmyEtAl2000} to fit this emission with
static gas distributions, as the gas pressure of the plasma is
insufficient, on its own, to drive a wind from the Milky Way.  Our
group asked whether cosmic-ray momentum and energy could be
communicated, via the ``Streaming Instability'' \citep{Wentzel1968,
KulsrudPearce1969, KulsrudCesarsky1971}, to the thermal gas, and hence
be used to launch a wind.  We were not the first to ask this question;
building on much past research into the power of cosmic rays to affect
gas dynamics and help drive a wind \citep[e.g.,][]{Ipavich1975,
Skilling1975, Bloemen1991, BrMcVo1991, BrMcVo1993,
BreitschwerdtSchmutzler1994, Zirakashvili1996, PtuskinEtAl1997,
ZirakashviliVoelk2006}, we assembled a 1D hydrodynamical model of a
cosmic-ray pressure and thermal-pressure driven wind.  The equations
for this wind model are reviewed in the Appendix.

The model of \pOne differed from previous models in that we
concentrated solely on understanding the enhanced X-ray emission
towards the inner Milky Way.  Taking the hypothesis that the emission
is centered on the Galactic Center \citep{AlmyEtAl2000}, we limited
the maximum radial extent of the wind to 4.5\kpc in Galactocentric
radius.  We assumed the wind was launched from the midplane of the
Galaxy; the wind then flows vertically within flowtubes with
cross-sectional area given by a simple analytic form with a
characteristic expansion height that is a free parameter \citep[as
in][]{BrMcVo1991}.  We set the initial cosmic-ray pressure and
magnetic field strength to the values inferred from synchrotron
observations at the Galaxy's midplane (F01), assuming a vertical
magnetic field.  The initial gas pressure and density at the Galactic
midplane, which are relatively poorly known, were left as free
parameters.  In early tests, we found that for fiducial values of all
these parameters, winds could not escape from the very inner portion
of the Galaxy ($R \la 1.5$\kpc), so we hypothesized an annular
geometry, with the wind only occupying a thick ring from 1.5\kpc to
4.5\kpc.  \citep[This inner region without a wind may correspond to
the transition from a wind to an X-ray emitting atmosphere of
gravitationally confined supernova-heated gas hypothesized in more
massive spheroids by][]{Binney2009}.

Values for the three free parameters (gas pressure \& density at the
midplane, and the scale-height of flowtube expansion) were then found
by fitting the predicted X-ray emission of the wind to the observed
emission in \textit{ROSAT}'s R4 and R5 bands; the wind was found to
provide a significantly improved fit (by a factor of two in $\chi^2$)
over the static polytrope of \citet{AlmyEtAl2000}.  In addition, the
gas pressure and density values that were found in this process were
plausible for the inner Galaxy (see Table~\ref{newWindParams}).
Finally, the power required to launch this wind was approximately a
factor of two higher than the estimated supernova power, which was
plausible given the many approximations of the theory, but possibly
indicated that further model refinement and/or constraints were
necessary.  Despite the clear simplicity of this approximate model,
such a cosmic-ray pressure- and thermal-gas pressure-driven wind
within our Galaxy seemed to fit the observations well.

X-ray observations alone do not uniquely determine the structure of
the wind; comparisons of the model to synchrotron data allow a natural
and complimentary check.  We outline the synchrotron-emission
calculation and compare the predicted emission to the observations in
\S\ref{synchEmission}.  The predicted synchrotron emission is larger
than what is observed, so we vary other aspects of the wind model in
\S\ref{newModelSetup}.  The resulting best-fit model provides nearly
as good a fit to the X-ray observations as the original model of \pOne
and is more consistent with other known aspects of galactic structure.

\section{Calculating the Synchrotron Emission}\label{synchEmission}

Given the above magnetic field and cosmic-ray density (defined by the
wind model presented in Paper I), the synchrotron intensity along a
line of sight can be integrated via equation (3.20) from
\citet{GinzburgSyrovatskii1965}:
\begin{equation}
I_{\nu} = \frac{\sqrt{3} {\rm e}^3}{{\rm m_e} {\rm c}^2} \int n(E_{\rm
  e}, {\bf r}, {\bf k})~B({\rm r}) \sin \theta \left(
  \frac{\nu}{\nu_{\rm c}} \int_{\nu/\nu_{\rm c}}^{\infty}
  K_{5/3}(\eta) d\eta \right) dE_{\rm e} dr, \label{mainSynchEq}
\end{equation}
where e is the electron charge, ${\rm m_e}$ is the electron mass, and
c is the speed of light. $n(E_{\rm e}, {\bf r}, {\bf k})$ is the
cosmic-ray electron density, which is a function of cosmic-ray
electron energy, $E_{\rm e}$, position, ${\bf r}$, and direction,
${\bf k}$.  $K_{5/3}$ is the Bessel function of the second kind,
$B({\bf r})$ is the magnetic field strength as a function of position,
$\theta$ is the angle between ${\bf B}$ and ${\bf k}$, $\nu$ is the
observation frequency, and $\nu_{\rm c}$ is defined by:
\begin{equation}
\nu_{\rm c} \equiv \frac{3 {\rm e} {\rm B_\perp}}{4 \pi {\rm m_e} {\rm
c}} \left( \frac{E_{\rm e}}{{\rm m_e} {\rm c}^2} \right)^2.
\end{equation}
For evaluating and integrating the full synchrotron intensity
equation, $n(E_{\rm e}, {\bf r}, {\bf k})$ is required throughout the
wind: we calculate $n(E_{\rm e}, {\bf r}, {\bf k})$ by assuming a
given initial electron distribution (\S\ref{initCReDistribution}), and
accounting for the integrated inverse-Compton, synchrotron, and
adiabatic losses (\S\ref{crCooling}) as the wind is launched from the
disk.  After we give the details for the calculation, below, we
briefly outline the convergence tests the code has passed
(\S\ref{testingSynch}), define the backgrounds (\S\ref{bkgds}) and
foregrounds (\S\ref{frgds}) that must be included, and show the
prediction for synchrotron emission from the wind in
\S\ref{originalModelComparison}.

\subsection{The Galactic Midplane Cosmic-Ray Electron Spectrum}\label{initCReDistribution}

The wind equations for this model consider both the thermal gas and
cosmic rays as fluids.  To model the synchrotron radiation of the
cosmic-ray electrons, we have to model in more detail the cosmic-ray
electron spectrum with energy.  To do this, we first take, as a
typical cosmic-ray electron spectrum, the electron spectrum observed
at the Galactic midplane at the Sun's position (F01).  As in F01, we
use the normalization from Figure 6 of \citet{Webber1983} where, at
2.3\gev, $dn/dE = 17.3\,{\rm m}^{-2}\,{\rm sec}^{-1}\,{\rm
ster}^{-1}\,{\gev/{\rm nucleon}}^{-1}$.  We then convert this result
from the Sun's position to a Galactocentric radius of $R = 3.5$\kpc by
multiplying that local cosmic-ray electron flux by a factor of 2.4, as
derived in Equation~11 of F01. This factor comes from fitting the
variation of synchrotron radiation in the Galactic midplane as
observed by \citet{BeuermannEtAl1985}.

For consistency, we also use the cosmic-ray electron-spectrum
power-law of $\gamma = 2.5$ from F01.  We assume that this cosmic-ray
electron spectrum is constant at the Galactic midplane, and constant
over time, evolving only with height in the wind due to cooling.  Of
course, if the CR e$^-$/p ratio changes towards the center of the
Galaxy from that observed locally, such a change would produce a
directly proportional change in the synchrotron intensity.  We point
out that a different input spectrum may be possible, as
\citet{StrongEtAl2004b} find that their Galactic propagation models
imply a different electron-injection spectrum from that used in \pOne
or in the present work.  Such uncertainties represent a large area of
parameter space that adds more complexity via another free parameter,
and so we choose instead to retain the present spectrum, based on
local measurements.  Our synchrotron results could easily be scaled to
lower or higher CR e$^{-}$ densities; as we will see in
\S\ref{newModelSetup}, this is not necessary to fit the observed
synchrotron emission.

We note briefly that the above procedure is similar to that we used to
estimate the cosmic-ray proton pressure at the base of the wind in
\pOne; as in F01, we have assumed that the local cosmic-ray proton to
cosmic-ray electron ratio applies also to the area at the base of this
wind.  This is mentioned further in \S\ref{improvedWindModel}.  The
electron-to-proton ratio is set to 1:50; one could assume
various ratios up to 1:100, perhaps, but assuming 1:50 should force
relatively strong constraints on the wind model when comparing to
radio synchrotron emission.


\subsection{Cooling the Cosmic-Ray Electrons}\label{crCooling}

To include the effects of inverse-Compton cooling on the cosmic-ray
electrons, we use the representation of the interstellar radiation
field (ISRF) that has been developed for use with the GALPROP code
\citep[available at \texttt{http://galprop.stanford.edu/};
see][]{StrongMoskalenko1998, StrongEtAl2000, StrongEtAl2004b,
PorterEtAl2005, PorterEtAl2008}\footnote{The ISRF data file is
\texttt{http://galprop.stanford.edu/FITS/
MilkyWay\_DR0.5\_DZ0.1\_DPHI10\_RMAX20\_ZMAX5
\_galprop\_format.fits.gz}, downloaded on September 23rd, 2008 from
\texttt{http://galprop.stanford.edu/web\_galprop/
galprop\_home.html}.}.  In their model, the ISRF is given as a
function of wavelength on a regularly spaced grid in Galactocentric
radius and height above the plane: the radial grid has 41 elements in
Galactocentric radius, starting at 0.25\kpc and spaced at 0.5\kpc
increments.  The height grid starts at $z=-5\kpc$ and is spaced in
increments of 0.1\kpc, in 101 steps, to $z=5\kpc$.  The ISRF photon
field is then converted to $f_\gamma(\omega)$, the background photon
distribution \citep[see Appendix C,][]{StrongMoskalenko1998} via the
definition:
\begin{equation}
U_{\rm ph} = {\rm m_e} {\rm c}^2 \int \omega^3 f_{\gamma}(\omega) d\omega,
\end{equation}
where $U_{\rm ph}$ is the energy density of the ISRF, $\omega$ is the
photon frequency, and $\gamma$ is the electron Lorentz factor.  With
the ISRF photon distribution, we can then write the equation for
inverse-Compton cooling \citep[see, again, Appendix C of][and
  references therein]{StrongMoskalenko1998} as:
\begin{equation}
\left(\frac{dE}{dt}\right)_{\rm IC} = \frac{\pi {\rm r_e}^2 {\rm m_e}
    {\rm c}^3}{2 \gamma^2 \beta} \int_0^\infty
    f_{\gamma}(\omega)[S(\gamma, \omega, k^+) - S(\gamma, \omega,
    k^-)],\label{icEq}
\end{equation}
where
\begin{eqnarray}
S(\gamma,\omega,k)&=&\omega
\left\{ \left(k+\frac{31}{6}+\frac{5}{k}+\frac{3}{2k^2}\right)\ln(2k+1)
-\frac{11}{6}k-\frac{3}{k}+ \right. \nonumber \\
&& \left. \hspace{5ex} \frac{1}{12(2k+1)}+\frac{1}{12(2k+1)^2}
+Li_2(-2k)
\right\} \nonumber \\
&&-\gamma
\left\{ \left(k+6+\frac{3}{k}\right)\ln(2k+1)
-\frac{11}{6}k+\frac{1}{4(2k+1)}- \right. \nonumber \\
& & \left. \hspace{5ex} \frac{1}{12(2k+1)^2}
+2Li_2(-2k)
\right\}\label{kleinNishinaS},
\end{eqnarray}
and where $Li_2(\cdots)$ denotes the dilogarithmic function \citep[the
  polylogarithm with n=2; see][]{AbramowitzStegun1972}, $\beta = v/c$
for the cosmic-ray electrons, and $k^{\pm} \equiv \omega \gamma(1 \pm
\beta)$.  For completeness, we note that the full application of this
formula to the production of $\gamma$-ray emission requires the
consideration of anisotropic scattering
\citep[see][]{MoskalenkoStrong2000}; in the present case of simple
cooling, where we need only the absolute power emitted, the
anisotropic nature of the radiation field is not important.


We also include the energy loss from synchrotron cooling:
\begin{equation}
\left( \frac{dE}{dt} \right)_{\rm synch} = -\frac{32}{9} \pi {\rm
  r_e}^2 {\rm c} U_{\rm B} \gamma^2 \beta^2.\label{synchEq}
\end{equation}
Adiabatic cooling is already included via the hydrodynamic cosmic-ray
model of the wind.  

In summary, then, the hydrodynamic wind model includes the adiabatic
cooling, and then a separate synchrotron-modeling program implements
the energy-dependent inverse-Compton (Eq.~\ref{icEq}) and synchrotron
(Eq.~\ref{synchEq}) cooling losses for the energy resolved electron
distribution.

To calculate the effect of these cooling terms, the $\Delta t$ for
each discrete vertical step of the wind is calculated from the
velocity profile, and then simply multiplied by the above cooling rate
at every step.  The inverse-Compton and synchrotron cooling do
substantially impact the energy density of the cosmic-ray electrons.
It is important to note, however, that neither process impacts the
cosmic-ray protons, which are the source of momentum for driving
winds; cosmic-ray electrons, with a number density only approximately
2\% that of the cosmic-ray protons, do not contribute significantly to
the cosmic-ray pressure.  So, in this case, when we consider
cosmic-ray electrons as a separate component, and cool those electrons
in an already defined wind model, that cooling does not impact the
wind model (such as the mass outflow rate and velocities achieved, for
instance).

Following the calculation of the local electron spectrum, the local
intensity is calculated and then integrated numerically on lines of
sight through the wind; synchrotron self-absorption does not play a
role in the Galaxy for $\nu \ga 50$\mhz, so it is neglected here.
This calculation is carried out over a grid in Galactic longitude \&
latitude.  We then average the emission over longitude (as done
previously for the comparison with the soft X-ray emission) and plot
it as a function of latitude to compare with observations.

How do the various cooling terms compare?  We have found that
inverse-Compton and synchrotron cooling of cosmic-ray electrons
dominate adiabatic cooling by a factor or $\sim 10$ near $z \sim
500\pc$, but that adiabatic cooling achieves parity around $z \sim
2\kpc$ above the Galactic midplane, reaching a maximum in relative
strength (2.5 times the total of inverse-Compton and synchrotron
cooling) at approximately 5\,kpc; above this height, adiabatic cooling
decreases again to become approximately equal to both inverse-Compton
and synchrotron cooling (we have calculated the above energy losses
over the entire electron-energy distribution that we consider, from
$\sim10^{-3}$\,\gev to $100$\,\gev).  The models also demonstrate that
inverse-Compton cooling is more important than synchrotron cooling by
approximately an order of magnitude near $R = 3.5\kpc$.  Also, we have
found that Klein-Nishina effects are significant at 408\mhz, with
total inverse-Compton cooling dropping to 65\% of that calculated from
the non-relativistic approximation.

\subsection{Testing the Synchrotron Code}\label{testingSynch}

Before running the full synchrotron calculation with cooling, we
developed an initial version of the code that assumed a constant
(non-cooling) cosmic-ray electron population with a power-law index of
2.7.  This code was checked against analytic estimates for the
synchrotron emission along various lines of sight, and was used to
check convergence of the flux integration with various line-of-sight
stepsizes; the algorithm reproduced the analytical results within
$\sim 5\%$.

We have also tested the synchrotron code by checking for convergence
in a wide range of parameters used within the code: for instance, we
have checked that the code converges when using a larger number of
electron bins in the CR-electron spectrum, smaller vertical step-sizes
in the cooling calculation, and higher resolution in all numerical
integrals.  All of the higher-resolution studies yield less than 5\%
differences in the predicted synchrotron emission.  This algorithm
also reproduces (within 3\%) the result of \citet{Ferriere2001} for the
synchrotron emission of the Galactic midplane.  Finally, the code can
also be run in a ``static halo'' mode without any cooling terms or
flow velocities; those runs reproduced the expected flat spectral
indices to approximately 1 part in $10^4$.

\subsection{Synchrotron Backgrounds}\label{bkgds}

Before comparing to observations, we must also account for the various
backgrounds in the data, adding those components to the wind model's
synchrotron emission.  As shown in \citet{ReichEtAl2004}, and in more
detail in \citet{ReichReich2008}, the observed brightness temperature
($T_{\rm OBS}$) can be understand as the sum:

\begin{equation}
T_{\rm OBS} = T_{\rm GAL} + T_{\rm CMB} + T_{\rm EXG} + T_{\rm OFF}
\end{equation}
where
\begin{eqnarray}
T_{\rm GAL} & = & {\rm intrinsic~Galactic~brightness~temperature}\\
T_{\rm CMB} & = & {\rm cosmic~microwave~background~=~2.73\K}\\
T_{\rm OFF} & = & {\rm zero-level~error} \\
T_{\rm EXG} & = & {\rm unresolved~extragalactic~sources}
\end{eqnarray}
We calculate the extragalactic contribution from the formula
\citep[see][and references therein]{ReichEtAl2004}:
\begin{equation}
T_{\rm EXG} = 30\K \left( \frac{\nu}{178\mhz} \right)^{-2.9}
\end{equation}
and take $T_{\rm OFF}$ from \citet{ReichEtAl2004}.

\subsection{A Strong Local Synchrotron Foreground?}\label{frgds}

After adding in all of these background components to the model, we
were intrigued to find a relatively constant level of high-latitude
(at $|b| \ga 40\degrees$) emission that remained unexplained, and
would not be explained by the wind emission, which would only be
important up to $|b| \sim 20\degrees$.  The Galactic synchrotron model
of \citet{BeuermannEtAl1985} also does not address this component;
they only fit the synchrotron emission at low latitudes ($b \la \pm
3\degrees$).  So, it seems that the nature and origin of this
high-latitude component are not well understood, and have not been
accounted for in the ``standard model'' of \citet{BeuermannEtAl1985}.
From the near constancy of the high-latitude emission, from the recent
evidence for a relatively local synchrotron emission component
\citep[][and W. Reich (personal communication)]{FleishmanTokarev1995,
RogerEtAl1999, Wolleben2007, SunEtAl2008}, and from our own (brief)
analysis of the 408\mhz to 1420\mhz spectral indices which seem to
indicate a temperature spectral index near $\beta \sim 2.75$, we
conservatively hypothesize that this high-latitude emission at $|b|
\ga 40\degrees$ is local, and therefore is a foreground Galactic
component that should be accounted for separately when comparing the
wind to the observed emission.  This is important, as the wind model
should not predict a larger synchrotron emission than is observed in
any part of the sky; the presence of a local synchrotron foreground
makes this, of course, more difficult for the model, which is
therefore even more restricted by the observations.

In addition, as we will show later, these foreground components are
also essential in comparing the synchrotron spectral index of models
to observations, and so highlight the importance for a more global
model of the Galactic synchrotron emission \citep[perhaps like that in
GALPROP, but including a more general wind model;
see][]{StrongEtAl2004b, SunEtAl2008, WaelkensEtAl2009}.  \textit{It is
also important to emphasize that if this emission is local, then the
synchrotron halo of the Milky Way, as inferred from
\citet{BeuermannEtAl1985}, is substantially overestimated} (W. Reich,
personal communication).

\subsection{Synchrotron Emission from the Wind Model of \pOne}\label{originalModelComparison}

\begin{figure}
\begin{center}
\includegraphics[angle=90,width=8.8cm]{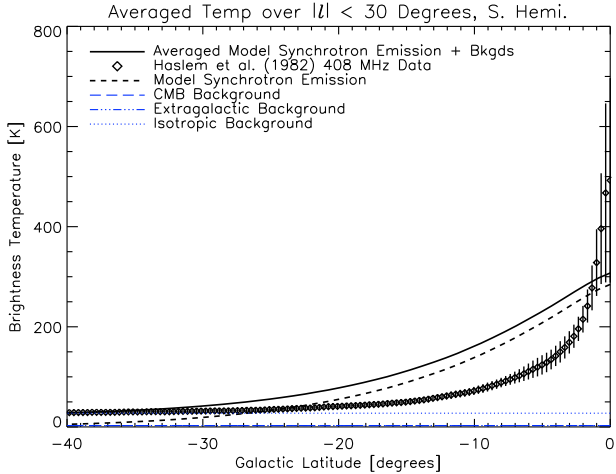}
\end{center}
\caption{The 408\mhz synchrotron brightness temperature predicted by
  the wind model presented in \pOne.  The predicted synchrotron
  brightness significantly overpredicts the observed emission by
  approximately a factor of two.
\label{synchTempsOriginalModel}}
\end{figure}

With all of these components added in, we calculate the absolute
intensity of synchrotron emission at 408\mhz with the original wind
model (as presented in \pOne) plus background and foreground emission.
We find that that model \textit{substantially} over-produces
synchrotron emission above the plane, as shown in
Figure~\ref{synchTempsOriginalModel}.  The wind does not dominate the
synchrotron emission in the disk, but, crucially, falls off too slowly
with height, violating observational constraints strongly near $b \sim
6\degrees$.  As such, the original model in \pOne is not in agreement
with the observations, and must be discarded as a ``best-fit'' to the
combined observations.  Clearly, either the assumed magnetic field is
too high, the wind volume is too large, or the cosmic-ray electron
density is too high to fit the observed synchrotron emission.

We now ask: is there a simple way for the model to be brought into
closer agreement with both the soft X-ray and synchrotron
observations, while at the same time, improving the basic assumptions
of the model of \pOne?

\section{Can an improved wind model fit both X-ray and Synchrotron
  Observations?}\label{newModelSetup}

We now consider two physically-motivated possibilities for updating
the previously assumed launch parameters of the wind as well as one
modification for making more robust comparisons between the model and
X-ray data.  These modifications are motivated by the overprediction
of synchrotron emission in the original model but also incorporate
current thinking about the structure of the inner Galaxy.  We then
check whether such changes may help bring a cosmic-ray and
thermally-driven wind model in agreement with the observations, and
consider the effect of these modifications on wind driving.

\subsection{A Wind From the ``Molecular Ring''}\label{thinnerWind}

As mentioned in the introduction, \pOne had included the assumption
that the wind was launched from a range of Galactocentric radii, from
1.5 to 4.5\kpc.  The inner radius was the minimum radius at which a
wind can be launched in the gravitational potential of the inner
Galaxy.  The observed edge of the soft X-ray emission at $l \sim
\pm30^\circ$ set the outer radius of the wind.  In the course of that
previous work, it was suggested by R. Benjamin (personal
communication) that the wind might more naturally be considered to be
launched from the ``5\kpc Molecular Ring''
\citep[e.g.,][]{JacksonEtAl2006}, which is much thinner, with a range
of Galactocentric radii of approximately $\Delta R \sim 1$ to 1.5\kpc.
This ``5\kpc Molecular Ring'' is home to the largest level of star
formation in our Galaxy, and so is a reasonable location for a wind
launched by hot gas and cosmic rays.  So, in this work, we attempt to
fit the X-ray emission with a wind that covers only 1\kpc on the
Galactic disk, from $R=3.5\kpc$ to $4.5\kpc$.  (This thinner wind may
hint at a more massive central Galactic bulge; it also increases the
importance of any X-ray emission from static gas in the bulge, which
we do not include in the model here.)

For future development of the model, we note that the ``Molecular
Ring'' is, of course, not observed to be uniform in azimuth, and so an
even more realistic assumption for launching would include a covering
fraction of the wind within this annulus.  This would require,
however, development of a more detailed model of the lateral force
balance as a function of height off the surface of the disk, which is
outside the scope of this paper.

\subsection{Lowering the Magnetic Field Strength}\label{lowerBField}

In the wind models presented in \pOne, we assumed the full magnetic
field strength at R = 3.5\kpc of $7.8\muG$ (F01).  It is certainly
not clear that the entire inferred field strength (in the plane of the
sky) would be present at the base of the wind, and evolve with the
wind to high latitudes.  In addition, if that magnetic field strength
were lower, the synchrotron emission would of course also decrease, as
$B^{(\gamma+1)/2} \sim B^{1.75}$ for $\gamma = 2.5$.  To test this in
the present work, we have decreased the magnetic field strength to
$5.2\muG$, or 66\% of the value used in \pOne; this is similar to
the local magnetic-field strength.  We are therefore assuming that
only this portion of the synchrotron-derived magnetic field at $R =
3.5\kpc$ is vertical field within the wind, and evolving with the
wind to mid- and high-latitudes.

In passing, we point out that the magnetic field strengths presented
in F01 are, in one sense, lower-limits to the field strength.  We have
found we are able to duplicate the results of the emissivity
calculation in F01 by using the formulae for homogeneous fields
perpendicular to the line of sight
\citep[see][]{GinzburgSyrovatskii1965}, so that the calculated
$7.8\muG$ magnetic field does not include the component of the field
along the line of sight.  On the other hand, it is important to note
that the emissivity used in F01 is actually the lowest of the
estimates of local emissivity in \citet{BeuermannEtAl1985} (as noted
in F01), so it is not straightforward to estimate the uncertainty in
the magnetic field towards the inner Galaxy.

In addition, it may be possible that the field in the wind is not
strictly vertical, but rather helical (P. Biermann, private
communication), as it may reasonably have a strong toroidal component
due to shear in the Galactic disk.  This would change the field
strength in the plane of the sky, and thereby affect the synchrotron
emission as a function of Galactic longitude.  For simplicity, we
neglect this effect in the current work.

\subsection{Limiting the Effects of Uncertain Absorption}\label{truncChi2}

In the original fits in \pOne, we found the best-fit to the X-ray
emission by calculating a $\chi^2$ value throughout the entire
latitude range of $b = -0\degrees$ to $-90\degrees$.  (We do not fit
the emission in the Northern Galactic Hemisphere because of the more
complex absorption and local emission sources, as discussed in \pOne.)
However, the fit to the X-ray emission is strongly affected by
absorption for $b \ga -10\degrees$ and is dominated by the constant
background components at $b \la -40\degrees$.  Intervening absorption
was, and is, included in the models, but the distance to those
absorbers is unknown, and so the absorption is calculated as a screen
of material located between the observer and wind.  This is perhaps a
reasonable first assumption, but we can minimize the impact of that
uncertainty in the fit by only calculating the $\chi^2$ value between
$-40\degrees < b < -10\degrees$, which we do for all of the following
models.  This also limits the impact of uncertainty in the stellar
contribution to X-ray emission at low-latitudes \citep{MasuiEtAl2009,
RevnivtsevEtAl2009}, although such emission is more significant at
higher energies.  (In practice, we find that these modifications do
not significantly change the parameters of the best-fit models in the
previous or current work.)

\subsection{Putting it All Together: an Improved Wind
  Model}\label{improvedWindModel}

Decreasing the wind model's footprint, so that it stretches only from
3.5 to 4.5\kpc on the disk, and attempting a smaller magnetic-field
strength ($B = 5.2\muG$) for the wind, we integrate the hydrodynamic
equations for a range of initial (at the base of the wind) thermal-gas
pressures, thermal gas densities, and $z_{\rm break}$ values to find
the best-fit model (using the same methods of \pOne; see the Appendix)
to reproduce the X-ray emission.  This search entailed two large-scale
searches through parameters space that integrated the hydrodynamic
equations of motion for 13,475 models each.  The first search spanned
two orders of magnitude in thermal-gas density at the base of the
wind, over one order of magnitude in thermal-gas pressures at the base
of the wind, and $z_{\rm break}$ values spanning a factor of two (in
35, 35, and 11 steps, respectively), centered roughly near the best
fit parameters from the wind model in \pOne.  The best-fit to the soft
X-ray emission from this scan was then input to a parameter-search of
another 13,475 models distributed over a slightly smaller range in
parameters (thermal-gas density ranging over a factor of ten,
thermal-gas pressure ranging over a factor of 2.25, and $z_{\rm
  break}$ ranging over a factor about 1.5), to zoom in near the best
fit. A final level of refinement in the parameter search was run to
scan 1331 models in those three parameters (only varying each
parameter by 40\%, 20\% and 20\% around the previous best-fit value
for $n_{\rm g,0}$, $P_{\rm g,0}$, and $z_{\rm break}$,
respectively).

\begin{figure}
\begin{center}
\includegraphics[angle=90, width=8.8cm]{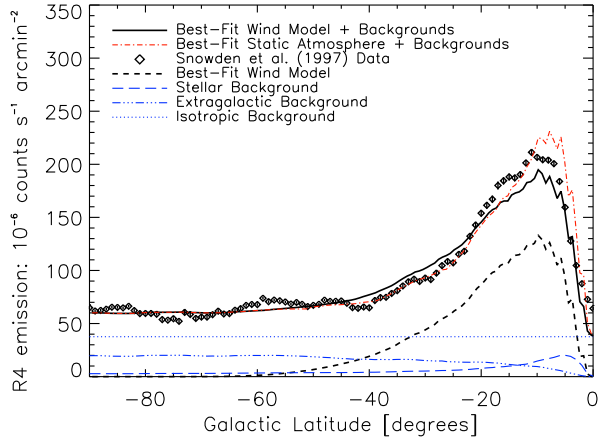}
\end{center}
\caption{The best-fit model for a Galactic wind that stretches only
  between 3.5 and 4.5\kpc in Galactocentric radius with $B=5.2\muG$;
  this comparison is for the ROSAT 'R4' band, centered roughly on
  0.65\kev.  The $\chi^2$ for this fit is 2335 for 87 degrees of
  freedom, yielding a $\chi_{\nu}^2$ of 26.8; this is slightly worse
  but comparable to the value of $\chi_{\nu}^2 = 19.0$ from the
  best-fit R4 model in \pOne. \label{r4Fit}}
\end{figure}

\begin{figure}
\begin{center}
\includegraphics[angle=90, width=8.8cm]{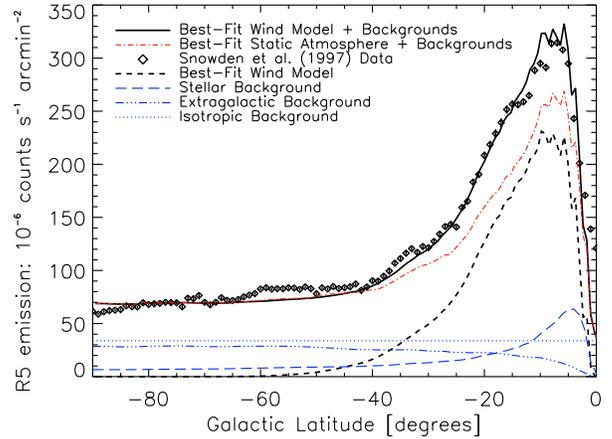}
\end{center}
\caption{As in Figure~\ref{r4Fit}, but for the ROSAT `R5' band that is
centered roughly on 0.85\kev.  The $\chi^2$ for this fit is 4522
with, again, 87 degrees of freedom; $\chi_{\nu}^2$ is then 52.0, which
again is only slightly worse fit than the $\chi_{\nu}^2 = 48.9$ fit
from \pOne. \label{r5Fit}}
\end{figure}

\begin{figure}
\begin{center}
\includegraphics[angle=-90, width=8.8cm]{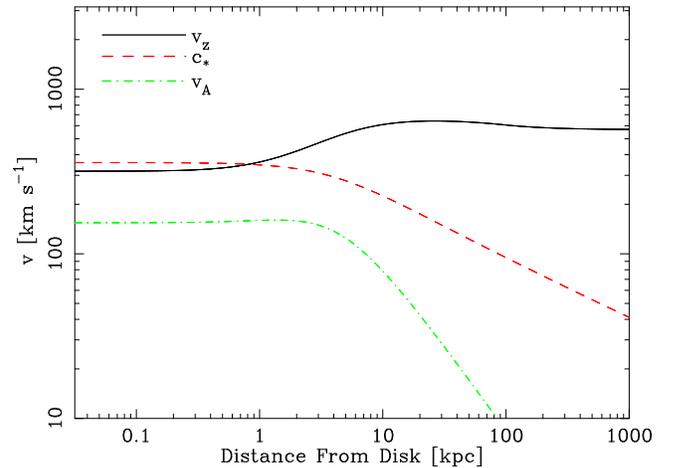}
\end{center}
\caption{Velocity vs height for the best-fit model (parameters given
  in Table~\ref{newWindParams}).  The solid line shows the thermal-gas
  velocity within the wind, the dashed, red line shows the composite
  sound speed as a function of height, and the dot-dashed, green line
  shows the variation in \alfven speed with height.  The increase of
  velocity (even through the critical point, where $v = c_*$) is
  fairly standard, although this wind exhibits a slight decrease in
  velocity at large distances, as the increased wind radius
  encompasses more and more of the Galaxy's mass.
  \label{velocityPlot}}
\end{figure}

\begin{figure}
\begin{center}
\includegraphics[angle=-90, width=8.8cm]{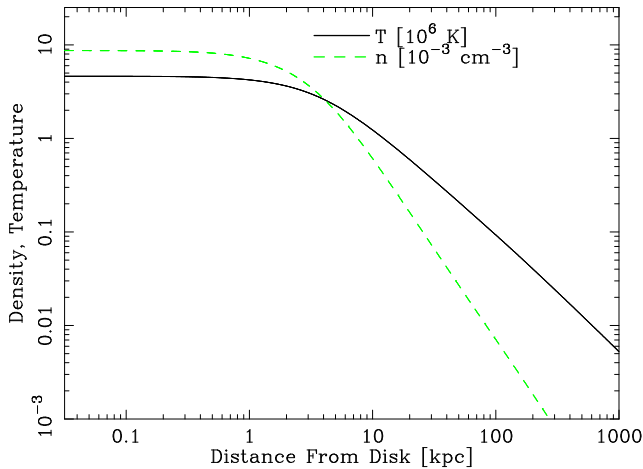}
\end{center}
\caption{The temperature and density in the thermal gas as a function
  of height in the wind.  The temperature is given by the solid, black
  line; as expected, as the wind accelerates and expands into the
  assumed flowtube, the temperature decreases.  The temperature
  does not decrease as quickly as without cosmic-ray heating,
  however.  Also in this plot, we show the particle-number density as
  a function of height with the dashed, green line.  \label{densityTemperaturePlot}}
\end{figure}

The new best-fit model, found by this procedure, is shown in
Figures~\ref{r4Fit} and \ref{r5Fit}, with the parameters for the fit
given in Table~\ref{newWindParams}.  We also show the velocity and
density profiles of this wind in Figures~\ref{velocityPlot} and
\ref{densityTemperaturePlot}.

This model still improves on the static model of \citet{AlmyEtAl2000}
with respect to $\chi^2$.  First, calculating $\chi^2$ over the entire
range of longitude ($-90\degrees < b < 0\degrees$, which is not the
restricted region based on which the model was selected), the new
model improves on the static polytrope's $\chi^2$ by a factor of 1.6
in the \textit{ROSAT} `R4' band and 2.1 in the `R5' band.
It is important to note that this fit is not as good as the original
fit from \pOne, however: the new model overpredicts the emission in
the X-ray `R4' band by 15\% at intermediate latitudes ($b \sim
-30^\circ$), and slightly overpredicts the emission in `R5' at low
latitude ($l \sim -10^\circ$)\footnote{It is important to stress that
this assumes that the wind model plus backgrounds explains all of the
X-ray emission observed towards the center of our Galaxy; it is very
conceivable (\textit{especially} now that the wind is much thinner
than in \pOne) that other components may help explain the emission
\citep[such as the local ISM, the Galactic bulge, or even a
bulge-related wind, see][]{TangEtAl2009}, and therefore the
requirements for the wind could be somewhat relaxed, and the strength
of emission in the wind therefore decreased.}.  (We note that this
best-fit model is defined as the model with lowest $\chi^2$,
calculated in the region from $-40\degrees < b < -10\degrees$ as defined
in \S\ref{truncChi2}, but this only changes the best-fit model
parameters by $\la 3\%$ from the fit calculated with $\chi^2$ defined
over the entire latitude range.)

How have the changes in $\Delta R$ and $B$ affected the wind model?
In order to fit the X-ray emission while being launched from a thinner
annulus, the gas density must increase at the base of the wind (here,
by $\sim 30\%$).  The gas pressure must increase as well, to keep the
wind at approximately the same temperature (which is constrained by
the relative emission in the `R4' and `R5' bands).  The gas pressure
has to increase still further for another reason, as well: as the
magnetic-field strength has decreased and the density has increased,
the power that is channeled from the cosmic rays to the gas also
decreases (the power goes as $v_{\rm A} \cdot \nabla P_{\rm cr}
\propto \frac{B}{\sqrt{\rho}} \cdot \nabla P_{\rm cr}$).  In the model
of \pOne, energy from cosmic-rays played a significant role in heating
the gas at mid-latitudes of $b \sim 15\degrees$.  But, as $B$
decreases and $\rho$ increases, the power transmitted from the
cosmic-ray component decreases, and the base gas temperature must
increase still further to compensate (see Eq.~\ref{gasPressureEq} of
the Appendix).  This leads to an increase in the gas
pressure at the base of the wind of approximately $104\%$.  These are
the dominant changes to the model as a result of the synchrotron
constraints on the wind model.

Turning to other parameters of the model, the lengthscale for flow
tube expansion ($z_{\rm break}$) has decreased by $\sim 25\%$ (this
parameter sets the scale-height for the expansion of the assumed flow
tubes, and so, in a sense, represents an important set of assumptions
about the magnetic geometry of the Galactic halo, which helps to
define the acceleration profile).  Meanwhile, the cosmic-ray pressure
at the Galactic midplane has remained constant, by assumption.  To
further compare models, the mass outflow rate in this wind is
$2.2\,M_{\odot}$/yr (very similar to the $2.1\,M_{\odot}$/yr for the
previous model), and the terminal velocity is $\sim 570\kms$,
significantly lower than the $760\kms$ for the wind of \pOne.
Overall, the model is somewhat similar to the past one in its ability
to fit the X-ray emission, and in its parameter values, with the
notable exception of strongly increased gas pressure.  It is also
important to point out that the initial velocity of gas in this simple
wind model is 252\kms; such a high initial velocity is the result of
our assumption that all of the energy is input as a delta function at
the Galactic midplane, and so strongly indicates the need for a more
detailed model of more gradual momentum and energy input to the wind.

\begin{figure}
\begin{center}
\includegraphics[angle=90,width=8.8cm]{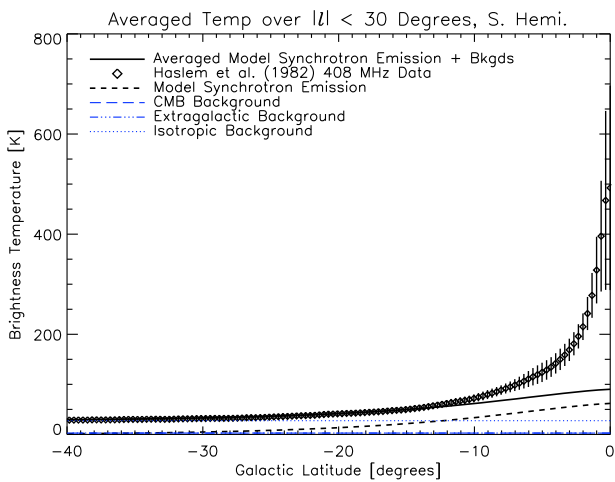}
\end{center}
\caption{The integrated 408\mhz synchrotron brightness temperature
  given by a wind that covers a smaller area on the Galactic disk than
  the model in \pOne; this wind extends only from 3.5\kpc to 4.5\kpc
  on the disk, and has $B=5.2\muG$. \label{synchTempsThinWind}}
\end{figure}

Given this best-fit model to the soft X-ray emission, we can compute
the synchrotron emission for the model.  We compare the resultant
brightness temperature to 408-MHz observations in
Figure~\ref{synchTempsThinWind}.  This figure shows the averaged
brightness temperature from the 408-MHz survey compared to the wind
model outlined above (when all background and foreground sources have
been added); both the data points and model curve are the result of an
average over $-30^{\circ} < l < 30^{\circ}$.  The wind model now
successfully fits the observed synchrotron emission at the
mid-latitudes of $-25^\circ < b < -10^\circ$.

\begin{figure}
\begin{center}
\includegraphics[angle=90,width=8.8cm]{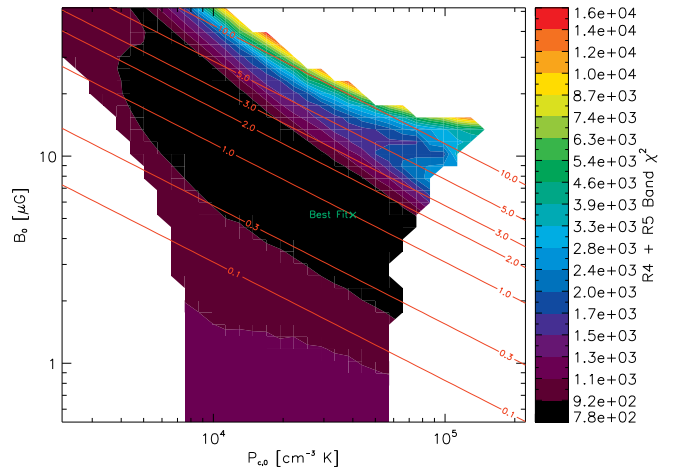}
\end{center}
\caption{The parameter space of the best-fitting model as a function
  of the magnetic-field strength at the base of the wind, $B_0$, and
  the cosmic-ray pressure at the base of the wind, $P_{\rm c,0}$.  The
  color-contour-delineated region denotes those winds which can be
  launched from the Galactic potential, and the color contours show
  the combined, total $\chi^2$ for fitting the X-ray emission in the
  \textit{ROSAT} `R4' and `R5' bands.  The labeled red lines indicate
  the trend of synchrotron emission, with the number giving the
  estimated level of synchrotron emission \textit{above} that observed
  in the best-fitting model.  One can see a range of models with radio
  emission of the level of that observed (the red line marked `1.0')
  with relatively low $\chi^2$ values. The best-fit model to the soft
  X-ray observations (with $P_{\rm c,0}$ fixed) is indicated by the
  green `x'.  \textbf{NB}: The red lines here give the synchrotron
  scaling of $I_{\nu} \propto n_{\rm cr,e} B^{1.75}$ appropriate for
  $\gamma = 2.5$, and \textbf{do not} represent the results of a set
  of synchrotron simulations, which would yield modified curves due to
  variations in wind acceleration profiles impacting the run of
  density with height.  For winds in this survey, the parameters
  $P_{g,0}$, $n_0$, $z_{\rm break}$ and $\alpha$ were held fixed.
  \label{killerPlot}}
\end{figure}

We point out that this model fits simply due to the decreased $\Delta
R$ and $B$; as such, the model presented here is \textit{not unique}:
we have essentially modified the model to include more physically
motivated launching conditions, finding a model that satisfies the
observational requirements.  There are, of course, a range of models
that might also fit the observations, given corresponding changes in
$B$ and $P_{\rm c,0}$: the range of possible models in this parameter
space are shown in Figure~\ref{killerPlot}.  This figure displays the
range of parameter values for $P_{\rm c,0}$ and $B$ for which winds
are successfully launched; the color contours show the region of
escaping winds with the colors encoding the $\chi^2$ value for the fit
to the two bands of soft X-ray emission, while the red lines show an
estimate of the factor by which models in that parameter space would
over- or under-predict the synchrotron emission: the red lines show
the increase in synchrotron emission by a simple scaling of the
best-fit synchrotron emission by $n_{\rm cr,e} \cdot B^{1.75}$.
\textit{This figure shows the range of magnetic-field strengths and
  cosmic-ray pressures that would approximately satisfy the
  synchrotron (with the red line labeled `1.0') and X-ray
  observations}.  The figure is also interesting in its own right, as
it shows the wide range of magnetic-field strengths that can help to
launch a wind.  The biggest difference between the various models over
the range of $B$ is that the models with larger $B$ can more easily
reproduce the large scale-height of soft X-ray emission due to
distributed heating by the cosmic-ray protons.

\begin{figure*}
\begin{center}
\includegraphics[angle=90,width=14cm]{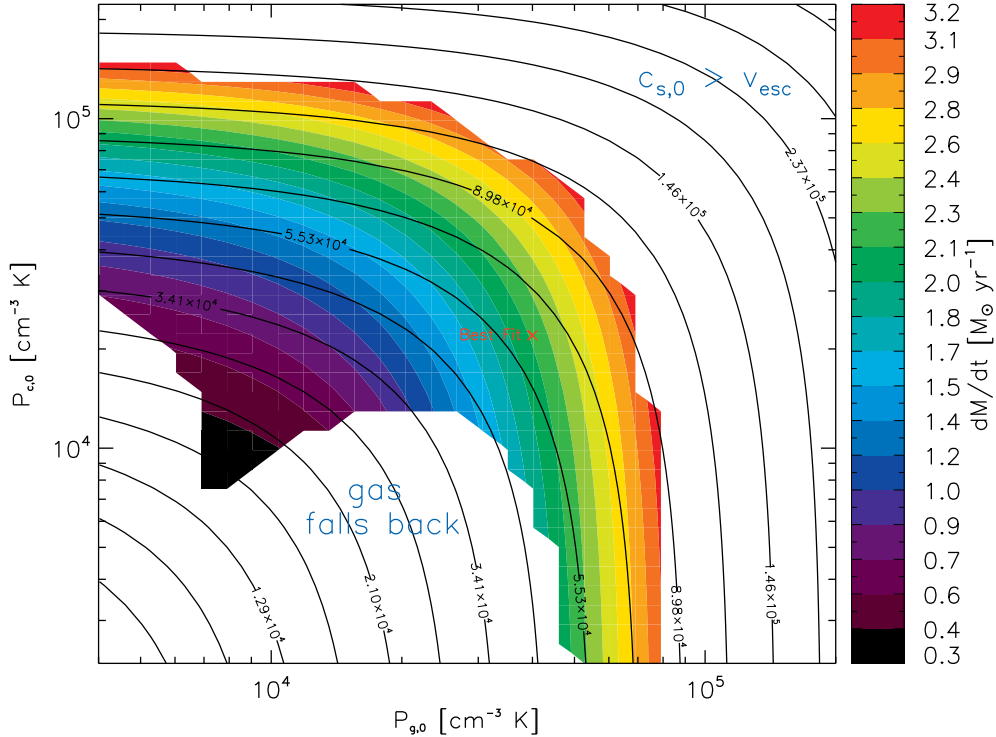}
\end{center}
\caption{The parameter space of successful outflows as a function of
  gas pressure at the base of the wind, $P_{\rm g,0}$, and cosmic-ray
  pressure at the base of the wind, $P_{\rm c,0}$.  The color contours
  outline the region of allowed winds, with the color code specifying
  the mass outflow rate in the wind.  However, in this figure, the
  black lines denote the total pressure in the wind (cosmic-ray
  pressure plus thermal pressure).  As expected, the mass outflow rate
  generally increases with total pressure in the wind.  This figure
  also shows the important role played by cosmic-ray pressure, though:
  cosmic-ray pressure above $P_{\rm c,0} \sim 10^5$\,cm$^{-3}$\,K
  result in successful winds for values of $P_{\rm g,0}$ that would
  ordinarily not allow outflows.  For winds in this survey, the
  parameters $n_{0}$, $B_0$, $z_{\rm break}$ and $\alpha$ were held
  fixed.
  \label{pc0vspg0}}
\end{figure*}

\begin{figure*}
\begin{center}
\includegraphics[angle=90,width=14cm]{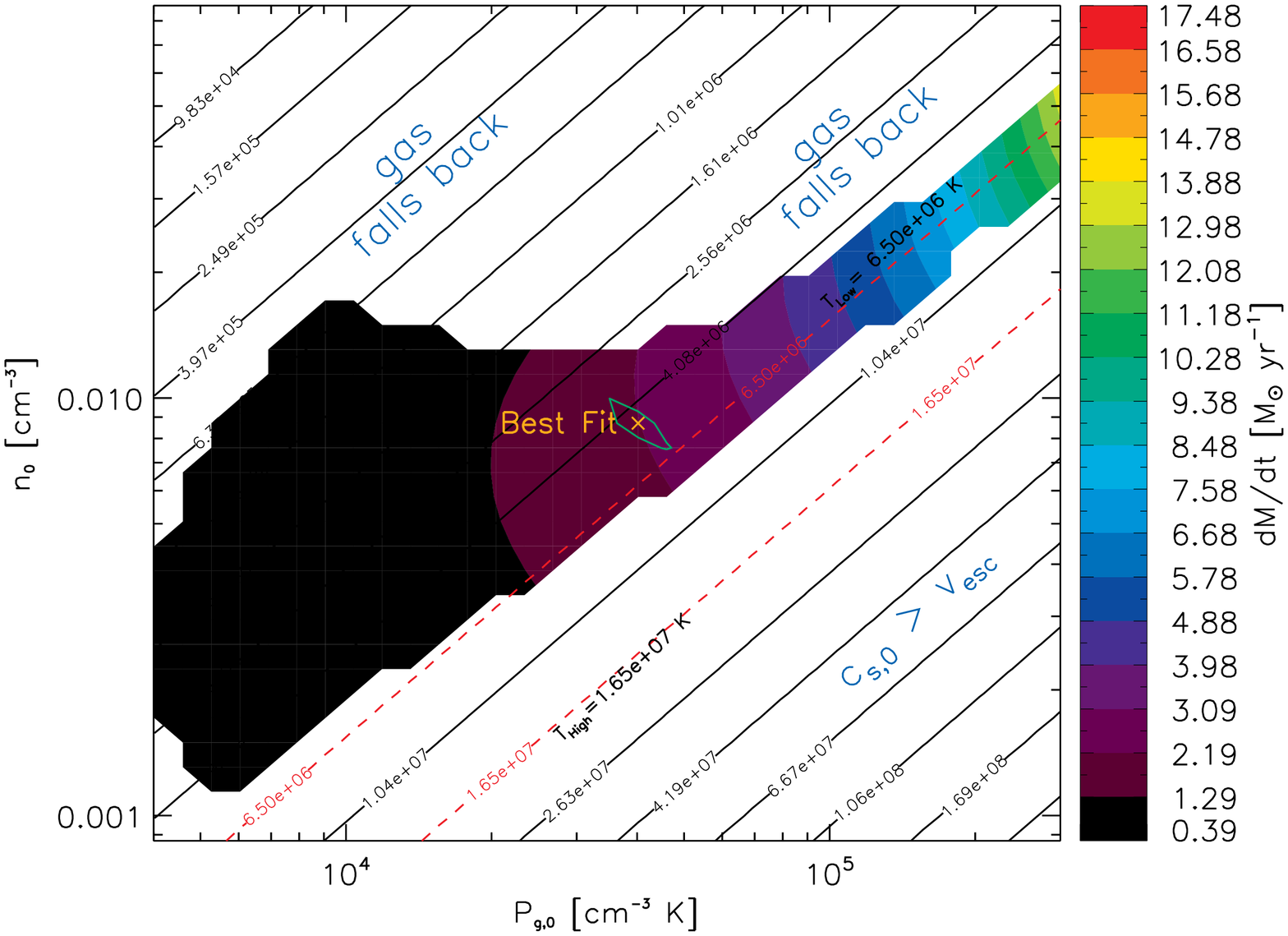}
\end{center}
\caption{The parameter space of allowed models as a function of gas
  pressure at the base of the wind, $P_{\rm g,0}$ and density, $n_0$.  The
  color contours show the range of models where plasma escapes the
  Galaxy, with the color corresponding to the mass outflow rate, given
  in the color bar on the right.  The solid lines are marked with the
  gas temperature along that line; the red-dashed line give the lower
  ($T_{\rm low}$) and upper ($T_{\rm high}$) limits for the gas
  temperature of a purely thermally-driven Galactic wind.  The best
  fit is shown with the yellow `x', surrounded by the region where
  $\chi^2 < 2 \cdot \chi^2_{\rm min}$, giving an indication of the
  region of minimally acceptable fits to the X-ray emission.  Compared
  to the similar figure in \pOne (that paper's Fig.~10), the region of
  allowed winds is smaller at higher pressure due to the relative
  dominance of gas pressure in this winds.  For winds in this survey,
  the parameters $P_{c,0}$, $B_0$, $z_{\rm break}$ and $\alpha$ were
  held fixed.
  \label{killerPlotOrig}}
\end{figure*}

Note that we have added a constant foreground component (as well as
other already known background sources, mentioned in \S\ref{bkgds}) to
our model in Figure~\ref{synchTempsThinWind}.  If the high-latitude
emission is not local, the constraints on $\Delta R$ and $B$ would be
relaxed, and larger $\Delta R$ and $B$ would be allowed.

\begin{deluxetable}{lrrl}
\tablecaption{Wind Parameters: New and Old\label{newWindParams}}
\tablehead{ & \colhead{New} & \colhead{Old} & \\ \colhead{Parameter} &
  \colhead{Value} & \colhead{Value} & \colhead{Fixed?}}
\startdata
$P_{\rm g,0}/k_{\rm B}$ & $4.0 \times 10^4$~K~cm$^{-3}$ & $2.0 \times
10^4$~K~cm$^{-3}$ &
Varied\\ 
$\rho_0$ & $9.0 \times 10^{-27}$~g\,cm$^{-3}$ & $7.1 \times 10^{-27}$~g\,cm$^{-3}$ & Varied  \\ %
$z_{\rm break}$ & $4.0$\kpc & $5.2$\kpc& Varied   \\ 
$R_0$ Range [Galactocentric] &  $3.5$ to $4.5$\kpc & $1.5$ to $4.5$\kpc& Fixed  \\
$P_{\rm c,0}/k_{\rm B}$ & $2.2 \times 10^4$~K~cm$^{-3}$ & $2.2 \times
10^4$~K~cm$^{-3}$ & Fixed\tablenotemark{a}  \\ 
$B_0$ & $5.2~\mu$G & $7.8~\mu$G & Fixed\tablenotemark{a}  \\
$\alpha$ & $2.0$ & $2.0$ & Fixed 
\enddata
\tablenotetext{a}{Varied in Figure~\ref{killerPlot} only, and not fit to the data.}
\end{deluxetable}

This fit assumes that all other parameters are constant, to attempt to
limit the number of variables.  In particular, the cosmic-ray pressure
at the base of the wind, $P_{\rm c,0}$, has not been allowed to vary
in any fits (although Fig.~\ref{killerPlot} displays the parameter
space of $P_{c,0}$ and $B_0$ to show how joint variations affect
synchrotron observations).  In Figure~\ref{pc0vspg0}, we show the
parameter space where $P_{\rm c,0}$ and $P_{\rm g,0}$ are allowed to
vary together; this plot shows the importance of adding cosmic-ray
pressure: the area of allowed winds (shown by the color contours)
increases markedly for $P_{\rm c,0} > 10^4$\,cm$^{-3}$\,K.  In that
regime, Galactic winds exist for lower thermal pressures than would
otherwise be allowed.  This plot also shows the best fit (from
Table~\ref{newWindParams}); however, interestingly, when $P_{\rm c,0}$
is allowed to `float', a fit is preferred where $P_{\rm c,0}$ and
$P_{\rm g,0}$ are approximately equal.  However, this results in a
cosmic-ray electron density that, given our stated assumptions about
the ratio of the cosmic-ray electron-to-proton fraction
(\S\ref{initCReDistribution}) and other parameters in the fit, would
violate the radio-synchrotron constrains.  Of course, if the
proton-to-electron ratio was 100:1 instead of 50:1, as assumed here,
that would relax this constraint on the wind greatly.

Finally, in Figure~\ref{killerPlotOrig}, we show the allowed winds as
a function of the wind parameters $n_0$ and $P_{g,0}$.  This figure is
much like the corresponding Figure~10 in \pOne, albeit with a
`slimmer' region of allowed winds due to the increased importance of
gas pressure vs. cosmic-ray pressure in the newer model (due to the
smaller magnetic-field strength).

\subsection{Another Diagnostic: the Wind's Brightness Temperature Spectral Index}\label{specIndex}

There is another possible constraint on the wind from synchrotron
observations: the temperature spectral index.  After much remarkable
work on survey and survey calibrations over the last few decades
\citep[e.g.,][]{HaslamEtAl1982, Reich1982, ReichReich1986,
ReichEtAl2001}, maps exist of the synchrotron spectral index over the
sky \citep{ReichReich1988a, ReichReich1988b, ReichEtAl2004}.  In
preparation for a new analysis of the spectral index \citep[some early
results have already appeared in][]{ReichEtAl2004}, we can make
predictions of the temperature spectral indices that would be produced
by the wind.  We calculate the brightness temperature spectral index
from predicted 408\mhz and 1420\mhz.  This spectral index is defined
as:
\begin{equation}
\beta = -\frac{\log(T(\nu_2)/T(\nu_1))}{\log(\nu_2/\nu_1)},
\end{equation}
where for this analysis, $\nu_1 = 408$\mhz and $\nu_2 = 1420$\mhz.  As
in the case of total intensity, we calculate this temperature spectral
index for the average emission in the model and in the data for $l =
0\degrees$, and over the range of latitude where the wind emission
seems to dominate the observed emission, $10\degrees < |b| < 25\degrees$
(and also, where discrete sources are much less numerous).  For a
similar region, \citet{ReichEtAl2004} found the temperature spectral
index $\beta \sim 2.65$ at $b \sim 30^\circ$ (see their Fig.~3; this
measurement includes all components along the line of sight).

\begin{figure}
\begin{center}
\includegraphics[width=8.8cm]{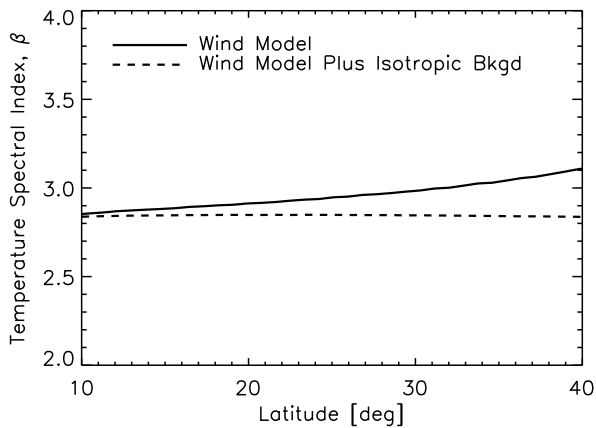}
\end{center}
\caption{The temperature spectral index, $\beta$, for the Galactic
  wind model along $l = 0\degrees$.  The solid line shows the trace of
  the spectral index for the wind model alone, while the dashed line
  gives the spectral index for the wind with a foreground emission
  component.\label{spectralIndexLong0}}
\end{figure}

The temperature spectral index for the best-fit wind model (alone) is
shown as the solid line in Figure~\ref{spectralIndexLong0}.  When we
consider the spectral index of the wind (without the foreground
emission component accounted for), the wind spectral index gradually
steepens, as expected for the cooling cosmic-ray electron population:
the wind model has $\beta \sim 2.85$ at $b = 10\degrees$, rising to
$\beta \sim 3.1$ by $b = 40^\circ$.  This is significantly steeper
than the spectral indices seen in \citet{ReichEtAl2004}, which
approximately span the range of 3.0 to 2.6, \textit{falling} with
height above the disk.

This source of ``flattening'' of the observed synchrotron spectral
index has been the source of some debate
\citep[e.g.,][]{ReichReich1988a, Pohl1990}.  In this debate, we would
like to emphasize the importance of \textit{multiple emission
components} that may lie along the line of sight \cite[as also
mentioned in][]{Pohl1990}.  Our model (when considered alone) clearly
shows a steepening of the spectral index with height, but when we
calculate the spectral index of the wind plus a constant (again,
perhaps local) foreground, we find a nearly constant spectral index;
this is shown by the dashed line in Figure~\ref{spectralIndexLong0}.
We note that this is the simplest foreground possible, but it is
sufficient to significantly change the run of spectral index with
height.

As such, until we have a more detailed understanding of background and
foreground synchrotron sources, it will be difficult to draw
definitive conclusions about any one component.  What is of course
needed here is an improved understanding of the nature of the various
components and how they work together, in order to constrain the
various physical processes and dynamics of our Galaxy
\citep{StrongEtAl2004b, Cox2005}.

\section{Results \& Discussion}\label{results}

We have continued to build on the results of previous work, and on our
own \pOne, by applying radio-synchrotron constraints derived from the
408\mhz all-sky survey of \citet[][]{HaslamEtAl1982} to our cosmic-ray
and thermal-gas pressure driven wind model.  The original model of
\pOne overpredicts mid-latitude synchrotron radiation (although it
would approximately reproduce the observations if a cosmic-ray
proton-to-electron ratio of 100:1 was adopted; we retain the locally
observed ratio of approximately 50:1 to place the most stringent
constraint on the wind model).  However, a wind model can be found that
is consistent with the observations and with more physically-plausible
launch conditions, such as a thinner wind with $\Delta R \sim 1\kpc$
and with smaller $B\sim5\muG$.

Both the trend of star formation with Galactocentric radius and
radio-synchrotron observations point to a wind that is launched from a
smaller ``footprint'' than hypothesized in \pOne.  This range of radii
seems plausible if we envision the wind as launched from within the
inner spiral arm, or the inner molecular ring, which has been observed
in this radius range.  We also found that a smaller magnetic field
strength was required: the vertical field in the wind can be decreased
from 7.8\muG to 5.2\muG and fit the observations.  The resulting wind
model predicts a level of synchrotron emission that agrees with the
level observed at high latitude, does not overpredict synchrotron
emission at low- to mid-latitudes ($b \sim -15\degrees$), and
reproduces the X-ray emission in the \textit{ROSAT} `R4' and `R5'
bands.  However, the fit in the `R4' band, while still improving on
previous static models, does not fit as well as model in \pOne.  We
have suggested that this may be due to the lack, in this model, of
X-ray emission from static gas.  Still, the constraint of the
synchrotron emission requires that the wind have a smaller radial
extent, and have a smaller magnetic field; the `best-fit' model, in
this paper, represents the best agreement between these two different
constraints (while keeping $P_{\rm c,0}$ constant).

While comparing the synchrotron observations to the wind model, we
have found that a significant fraction of high-latitude ($b <
-40\degrees$) synchrotron radiation is consistent with a local source
rather than Galactic-halo emission.  This is in agreement with other
work \citep[see, e.g.][and references therein]{SunEtAl2008}, and hints
that the synchrotron halo model of \citet{BeuermannEtAl1985} is
perhaps a significant over-estimate of the actual Galactic halo.  As
the model of \citet{BeuermannEtAl1985} is still in widespread use,
this seems an important point to stress.  We include this local
emission to again strongly constrain this Galactic wind model, to see
if it can survive.  It appears that it can.

Choosing this wind model has other implications for the effect of a
wind on the Galactic central region.  We wish to point out, however,
that we do not consider the diffusion of electrons throughout the
Galaxy; only the vertical advection of cosmic rays within a small
range of Galactocentric radii is considered.  Given this small range
of launching radii, and since the radial diffusion rate of cosmic-rays
is much less than the vertical advection rate, the present wind model
would not advect cosmic rays from a widespread range of Galactocentric
radii.  Indeed, we envision that, since cosmic rays stream mostly
along magnetic field lines, this wind's advection of cosmic rays would
not impact the cosmic-ray distribution in the solar neighborhood.

This advection of cosmic rays from the Galactic midplane could have
other effects, though: in \pOne, it was hypothesized that such a wind
could help evacuate cosmic rays from the central Galactic disk where
$\gamma$-ray observations seem to indicate that cosmic-ray protons are
not as abundant as the apparent supernova rate in that region of the
Galaxy would suggest \citep{Bloemen1989, BloemenEtAl1993, BrDoVo2002}.
This type of wind and its effect on the cosmic-ray population has been
studied in more detail by \citet{GedeBo09}.  The relative dearth of
cosmic rays towards the center of the Galaxy might then require an
additional component, such as a significant change in the $W_{\rm
  CO}$-to-$N(H_2)$ factor, as suggested by \citet{StrongEtAl2004a}.
This model will also require more localized supernova power to launch
the wind.  In the wind model presented in \pOne, the wind required
approximately 2.1 times the supernova power that is observationally
inferred in the radial range beneath it ($R = 1.5$ to $4.5\kpc$),
using the estimate of the supernova rate as a function of $R$ in F01.
Using the same method, the wind model presented in this paper requires
approximately 2.7 times the supernova power in the $\Delta R=1\kpc$
range of radii that it is launched from, assuming that each supernova
produces $10^{51}$\ergsPerSec \citep[using instead the supernova rates
  vs $R$ of][the wind requires 1.9 times the estimated supernova
  power]{McKeeWilliams1997}.  While, as mentioned in \pOne, thermal
conduction and clumping of the outflowing gas may help reduce this
estimate, this calculation shows that the synchrotron data has perhaps
pushed the simple wind model to the limit of its applicability to the
Galaxy.  However, this estimate is also very dependent on the radial
distribution of supernovae; if uncertainties in the distances to
supernovae are artificially spreading out the distribution of
supernovae, then the power in the Molecular Ring may be
underestimated.  All of these considerations point to the need for not
only a more complete model but also for continued comparisons of the
model with other observations.

Overall, this paper has shown that the wind model can satisfy both the
synchrotron observations and can be made more compatible with what is
already known of Galactic structure. This work satisfies those
observations and constraints by limiting the radial extent of the wind
and magnetic field strength within the wind.  The most direct way to
continue testing the model is to produce $\gamma$-ray predictions, and
compare those predictions with \textit{Fermi}/LAT observations of the
diffuse Galactic $\gamma$-ray emission.

\section{Acknowledgments}

The authors would like to thank T. Porter, A. Strong, and their
collaborators for their work on their representation of the
Interstellar Radiation Field, and for making their map of the ISRF
available online.  The authors also would like to thank Bob Benjamin,
Dieter Breitschwerdt, Katia Ferriere, Bryan Gaensler, Amanda Kepley,
Dan McCammon, Justin Morgan, Peter Biermann, Wolfgang Reich, Blair
Savage, and Eric Wilcots for questions and helpful conversation.  The
authors also thank the referee for comments that improved the paper.

Portions of the analysis presented here made use of the Perl Data
Language (PDL) developed by K. Glazebrook, J. Brinchmann, J. Cerney,
C. DeForest, D. Hunt, T. Jenness, T. Luka, R. Schwebel, and C. Soeller
and can be obtained from http://pdl.perl.org. PDL provides a
high-level numerical functionality for the Perl scripting language
\citep{GlazebrookEconomou1997}.

This work was supported by NSF AST-0507367, NSF AST-0907837, and NSF
PHY-0215581 \& NSF PHY-0821899 (to the Center for Magnetic
Self-Organization in Laboratory and Astrophysical Plasmas).  This
research has made use of NASA's Astrophysics Data System.

\appendix
\section{Wind Model Equations}\label{windEquations}

For completeness, we review here the equations for the cosmic-ray and
thermal-gas pressure driven wind from \citet{EverettEtAl2008}; please
refer to that paper for further details.

Our wind model builds on the equations first presented by
\citet{BrMcVo1991} and \citet{BrMcVo1993}.  This model is a 1D,
semi-analytic model that treats both the thermal gas and cosmic-ray
components as fluids.  The development of the outflow with height is
governed by the equation of mass conservation (assuming no mass is
added as the wind flows out of the Galactic plane):
\begin{equation}
\frac{d}{ds}(\rho v A) = 0
\end{equation}
where $\rho$ is the gas mass density, $z$ is the height above the
Galactic plane, and $A$ is the cross sectional area of the wind.
Since this is a 1D system of equations, we must prescribe the
cross-sectional area as a function of height:
\begin{equation}
A(z) = A_0 \left[ 1 + \left(\frac{z}{z_{\rm break}} \right)^\alpha \right]
\end{equation}
where $A_0$ is cross-sectional area of the flowtube at the Galactic
midplane, and $\alpha$ is the power-law governing the divergence of
the flowtube; the flowtube has a roughly constant cross section $A
\sim A_0$ until approximately the height $z_{\rm break}$, where the
area increases as $z^\alpha$.  We chose $\alpha = 2$ to mimic a
spherical divergence with height above $z_{\rm break}$ and leave
$z_{\rm break}$ as a parameter in our fitting process, as its value is
not known beforehand.

The thermal-gas and cosmic-ray pressure change with height via the
relations:
\begin{eqnarray}
\frac{dP_{\rm g}}{dz}& = &\left( c_{\rm g}^2 - 
               \gamma_{\rm c} (\gamma_{\rm g}-1) \frac{P_{\rm c}}{\rho} \frac{1}{M_{\rm A}}
               \frac{M_{\rm A} + \frac{1}{2}}{M_{\rm A} + 1} \right) \frac{d\rho}{dz},
               \label{gasPressureEq} \\
{\rm and}~~\frac{dP_{\rm c}}{dz} & = & \frac{\gamma_{\rm c} P_{\rm c}}{\rho} \frac{M_{\rm A} +
               \frac{1}{2}}{M_{\rm A} + 1} \frac{d\rho}{dz},
               \label{crPressureEq}
\end{eqnarray}
where $P_{\rm g}$ and $P_{\rm c}$ are the thermal-gas and cosmic-ray
pressures, $c_{\rm g}$ is the speed of sound, $\gamma_{\rm g}$ and
$\gamma_{\rm c}$ are the polytropic indices for the thermal gas and
cosmic rays (set to $5/3$ and $4/3$, respectively), and $M_{\rm A} =
v/v_{\rm A}$ is the \alfven Mach number for the gas ($v$ is the
velocity of the gas and $v_{\rm A}$ is the \alfven speed).

The magnetic-field strength evolves with height by requiring that the
magnetic flux remains constant; hence, $B(z) A(z) = {\rm [constant]}$.
We note that the magnetic field is important here not as a direct
source of pressure or tension, but as a ``conduit'' of sorts, through
which the cosmic rays communicate momentum and energy to the thermal
gas.  This is why we require a vertical magnetic field; a
magnetic-field component must lie along the direction of the
cosmic-ray pressure gradient in order to excite the streaming
instability \citep[e.g.,][]{Wentzel1968, KulsrudPearce1969,
BrMcVo1991}.

The terms in these equations (Eq.~\ref{gasPressureEq} and
\ref{crPressureEq}) can be understood as follows.  The first term in
Equation~\ref{gasPressureEq} describes how the gas pressure changes
due to expansion and acceleration of the gas in the flowtube.  The
second term in Equation~\ref{gasPressureEq} couples the cosmic-ray
pressure to the gas pressure; this term represents the damping of
cosmic-ray generated \alfven waves which then heat the gas and help
drive the wind.  This type of immediate wave damping is important as
rapid wave-damping mechanisms are known to occur in the ISM, and this
also limits the growth of the \alfven waves such that the perturbation
in magnetic-field strength remains below the background field
strength.  This idea was first introduced (but not yet widely used,
since the \alfven wave growth was small below the critical point) in
\citet{BrMcVo1991} and was also used in \citet{Zirakashvili1996} and
\citet{PtuskinEtAl1997}; we discuss this in more detail in \S2 of
\pOne.  Finally, Equation~\ref{crPressureEq} describes the loss in
cosmic-ray pressure due to adiabatic expansion, wave momentum
transport to the gas, and to wave generation by the cosmic rays.

The velocity of gas in the wind is the derived by solving the wind
equation, which we write as:
\begin{equation}
\rho v \frac{dv}{dz} + c_*^2 \frac{d\rho}{dz} = -\rho g.
\end{equation}
This equation of motion is much like the equation of motion for
models that rely only on thermal-gas pressure, except that instead of a
gas sound speed, $c_{\rm g}$, we use $c_*$, which gives a ``composite
sound speed'' \citep[see][]{BrMcVo1991}, defined as:
\begin{equation}
c_*^2 = \frac{d(P_{\rm g} + P_{\rm c})}{d\rho}
\end{equation}
where $P_{\rm g}$ and $P_{\rm c}$ were given in
Equations~\ref{gasPressureEq} and \ref{crPressureEq}.

To solve for the functions $v(z)$, $\rho(z)$, $P_{\rm g}(z)$, $P_{\rm
c}(z)$ and $B(z)$, we must integrate the above equations from the
Galactic midplane to a distance far above the galactic midplane,
correctly threading the critical point that occurs when $v = c_*$.
When starting the integrations from the midplane, we set (from other
models and observations) and fix the value of the midplane cosmic-ray
pressure, the magnetic field strength, and the power-law $\alpha$ that
governs the opening of the flowtube cross-sectional area (see
Table~\ref{newWindParams} for values).  We do not have strong
constraints on the values of the midplane thermal-gas pressure, the
hot-gas midplane density, and the height where the flowtube starts
to rapidly expand ($z_{\rm break}$) so we leave those as free
parameters which are then constrained by the soft X-ray and radio
observations.  These parameters are also considered in some detail in
\pOne and in \S\ref{windReview}.

\bibliography{ms}

\end{document}